\documentclass[prl,aps,twocolumn,showpacs,preprintnumbers,amsmath,amssymb]{revtex4}
\usepackage{dcolumn}
\usepackage{graphicx}
\usepackage{color}

\begin{document}

%\preprint{APS/123-QED}
\title{The Insulating Nature of Na$_{2}$IrO$_{3}$: Mott-type or Slater-type?}
\author{Minjae Kim}
\thanks{Present address: Coll\`{e}ge de France, 75005 Paris, France}
\author{Beom Hyun Kim}
\thanks{Present address: RIKEN, Saitama 351-0198, Japan}
\author{B. I. Min}
\email{bimin@postech.ac.kr}
\affiliation{Department of Physics, PCTP,
        Pohang University of Science and Technology, Pohang, 790-784, Korea}
\date{\today}

\begin{abstract}
We have investigated temperature-dependent electronic structures
of Na$_{2}$IrO$_{3}$ to unravel its insulating nature.
Employing the combined scheme of the density-functional theory (DFT) and
the dynamical mean-field theory (DMFT),
we have shown that the insulating state persists even above
the N\'{e}el temperature ($T_{N}$), which reveals that Na$_{2}$IrO$_{3}$ is
classified into a Mott-type insulator.
The measured photoemission spectrum in the paramagnetic (PM)  state is well described by
the electronic structure obtained from the DFT+DMFT
for the insulating state above $T_{N}$.
The analysis of optical conductivity, however, suggests that
the non-local correlation effect is also important in Na$_{2}$IrO$_{3}$.
Therefore, Na$_{2}$IrO$_{3}$ is not to be a standard Mott insulator in that
the extended nature and the non-local correlation effect of Ir 5$d$ electrons
are important as well in describing its electronic and magnetic properties.
\end{abstract}


\maketitle
%==============================================================================
%\section{Introduction}
Identifying the insulating nature of transition metal oxides
has been a central and long-standing subject in modern condensed matter physics.\cite{Imada}
Recent attention has been paid to 5$d$ transition metal oxides,
Sr$_{2}$IrO$_{4}$ and Na$_{2}$IrO$_{3}$, whether they belong to Mott-type or Slater-type
insulators.
The strong spin-orbit coupling (SOC) and the rather weak Coulomb correlation of
5$d$ electrons are known to be two essential ingredients in
determining the ground state physics of Sr$_{2}$IrO$_{4}$
and Na$_{2}$IrO$_{3}$.\cite{BJKim1,BJKim2,HJin09,Jackeli,Martins,Arita,SKChoi, Fujiyama12,
BHKim12,Comin,Singh,Mazin12,Chaloupka,Zhang,HSKim13,CHSohn13,Mazin13,Foyevtsova,BHKim14,HJKim14}
Despite intense studies on the role of interplay between Coulomb interaction and SOC,
however, there has been no consensus on the nature of their insulating states yet.
For example, on the insulating nature of Sr$_{2}$IrO$_{4}$,
there exist two contradictory reports,
Mott insulator\cite{Martins} vs. Slater insulator.\cite{Arita}
A marginal Mott insulating state was also proposed,
in which the insulating state above the N\'{e}el temperature ($T_{N}$=240 K)
was attributed to the presence of short range antiferromagnetic (AFM)
correlation.\cite{Fujiyama12,Zhang}
For Na$_{2}$IrO$_{3}$ too, which is a system of our present interest,
there have been debates on its insulating
nature.\cite{Chaloupka,Mazin12,Comin,Singh,CHSohn13,Mazin13,Foyevtsova,BHKim14,HJKim14}

Na$_{2}$IrO$_{3}$ exhibits insulating state
at room temperature ($T$), well above $T_{N}$=15 K.\cite{Comin,Singh}
The paramagnetic (PM) state with Curie-Weiss susceptibility behavior was
confirmed up to $T=500$ K.\cite{Singh}
The zigzag-type AFM ordering occurs below $T_{N}$.\cite{SKChoi,Ye}
To explain the AFM insulating ground state of Na$_{2}$IrO$_{3}$,
both the Mott-type\cite{Comin,CHSohn13} and Slater-type\cite{Mazin12,Mazin13,Foyevtsova,HJKim14}
mechanisms of the metal-insulator transition were invoked.
In the former, the good basis to describe the local electronic structure
is derived by consideration of the on-site atomic SOC and the crystal field.
This basis is usually very close to the relativistic
$J_{eff}$=1/2 orbital.\cite{Jackeli,HSKim13,BHKim14,Trousselet,Gretarsson}
Superexchange-based formalism is used to describe
the zigzag-type AFM ordering.\cite{Chaloupka,Yamaji}
The observed PM insulating state well above $T_{N}$ supports
the Mott-type mechanism.\cite{Comin}
In the Mott-type mechanism, however, very long range magnetic interaction
or very high energy excitation is required to describe
the zigzag-type AFM ordering, which raises the question on the validity
of superexchange-based formalism.\cite{Foyevtsova}
On the other hand, in the Slater-type mechanism, the quasi-molecular orbital (QMO)
has been suggested as a good basis with consideration of the
strong anisotropic hybridization between 5$d$ orbitals 
of neighboring Ir atoms.\cite{Mazin12} The zigzag-type AFM ordering
occurs as a consequence of the energy gain from the gap opening
at the zone boundary.\cite{Mazin13,HJKim14}
The Coulomb correlation of 5$d$ electrons is considered
just to enhance the band gap.
Larger extension of 5$d$ orbital and multi-peak features in
photoemission spectrum (PES) and optical conductivity
seem to support the Slater-type mechanism.
However, in the Slater-type mechanism,
it is hard to explain the observed PM insulating state well above $T_{N}$.
Therefore, the issue is whether the local AFM ordering is essential
in describing the insulating state of Na$_{2}$IrO$_{3}$ or not.

In this letter, we have investigated $T$-dependent
electronic structures and magnetic properties
of Na$_{2}$IrO$_{3}$, using the combined scheme
of the density functional theory (DFT) and
the dynamical mean-field theory (DMFT),\cite{Georges,Kotliar}
and unraveled its insulating nature.
We have shown that Na$_{2}$IrO$_{3}$ at room $T$
has the electronic structure of the PM insulating state,
revealing that Na$_{2}$IrO$_{3}$ is a Mott-type insulator.
Highly incoherent PES for Na$_{2}$IrO$_{3}$ in the PM state is well described
by the DMFT incorporating the local dynamical correlation.
However, we have also found evidences that Na$_{2}$IrO$_{3}$ deviates
from standard Mott insulators due to extended nature of Ir 5$d$ orbitals.
The onset of the zigzag-type AFM ordering induces the significant
redistribution of charge and spin densities with respect to those of $J_{eff}$=1/2 orbital.
This feature implies that, even though Na$_{2}$IrO$_{3}$ is
classified into a Mott-type insulator, the superexchange-based theory
assuming the rigid charge degree of freedom
needs to be refined to elucidate the magnetic ordering in Na$_{2}$IrO$_{3}$.
Moreover, optical conductivity, which corresponds to two-particle property,
is not well described by the DMFT incorporating only the local dynamical correlation.
Thus, the additional non-local spatial electronic correlation is expected to be important
to describe optical conductivity in Na$_{2}$IrO$_{3}$.

%%%%%%%%%%%%%%%%%%%%%%%%%%%%%%%%%%%%%%%%%%%%%%%%%%%%%%%%%%%%%%%%
\begin{figure}[t]
\includegraphics[width=8cm]{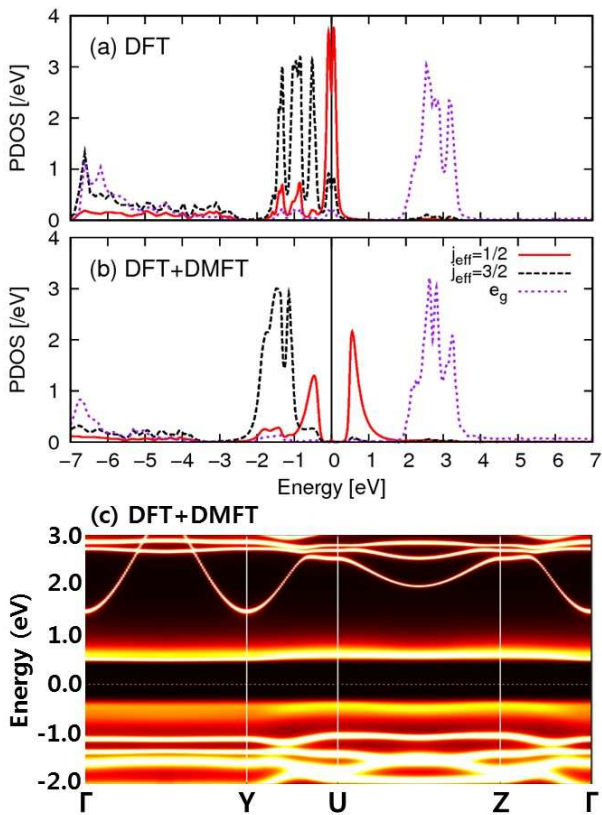}
\caption{(Color online)
Ir 5$d$ partial density of states (PDOS) of Na$_{2}$IrO$_{3}$
from the DFT (a), and the DFT+DMFT at $T=300$ K (b).
Red solid, black dashed, and purple dotted lines represent
$j_{eff}$=1/2, $j_{eff}$=3/2, and e$_{g}$ projected PDOSs, respectively.
See the text for numerical $j_{eff}$=1/2, $j_{eff}$=3/2, and e$_{g}$ basis.
(c) The spectral function from the DFT+DMFT at $T=300$ K.
High symmetry points of of Brillouin zone are
presented in the Supplemental Material.\cite{Supp}
}
\label{fig1}
\end{figure}
%%%%%%%%%%%%%%%%%%%%%%%%%%%%%%%%%%%%%%%%%%%%%%%%%%%%%%%%%%%%%%%

We have performed the fully charge self-consistent DFT+DMFT calculations
based on the projection-embedding scheme.\cite{Haule1,Haule2}.
Correlated Ir 5$d$ electrons were treated dynamically by the
DMFT local self-energy, while $s$ and $p$ electrons
were treated on the DFT level.
The SOC in the Ir 5$d$ orbital is included in all the calculations.
For Coulomb interaction parameters, $U$ and $J$,
we employed $U$=3.5 eV and $J$=0.8 eV for the DFT+DMFT.
These parameters yield the band gap and PES spectrum
in good agreement with the experiment.
The DFT part calculation was done by using
the full-potential linearized augmented plane wave
(FLAPW) band method.\cite{FLAPW,Blaha}
For the DFT+DMFT calculation, $T=300$ K was chosen, and the PM state is assumed.
We adopted experimental crystal structure of Na$_{2}$IrO$_{3}$.\cite{SKChoi}
We used numerical $j_{eff}$=1/2, $j_{eff}$=3/2, and e$_{g}$ bases
for Ir 5$d$ electrons in the DFT+DMFT,
which diagonalize the infinite frequency hybridization
function $\Delta(\omega=\infty)$ in the local impurity Green's function
of the DMFT result.
As shown below, these $j_{eff}$ bases are consistent with
relativistic $J_{eff}$ projectors for t$_{2g}$ orbitals
in the cubic symmetry.\cite{CHSohn13,HSKim13}
In comparison, we have also performed the DFT and DFT+$U$ calculations.\cite{Anisimov}
For the DFT+$U$, we used $U=2.75$ eV and $J=0.6$ eV,
following the existing calculation.\cite{Li}
We assumed, for the magnetic structure, the observed zigzag-type AFM ordering.\cite{SKChoi,Ye}
See the Supplemental Material (SM) for computational details.\cite{Supp}

%%%%%%%%%%%%%%%%%%%%%%%%%%%%%%%%%%%%%%%%%%%%%%%%%%%%%%%%%%%%%%%%
\begin{figure}[t]
\includegraphics[width=8cm]{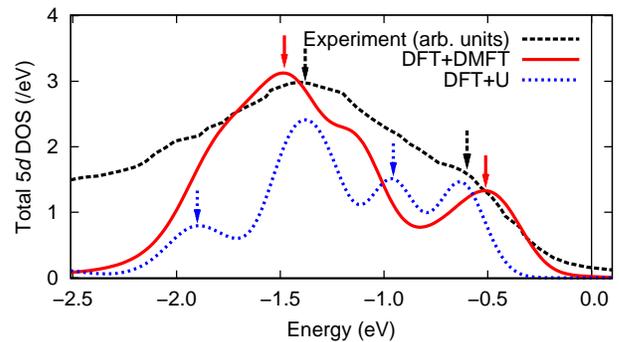}
\caption{(Color online)
Ir 5$d$ PES at $T=130$ K (black dashed)\cite{Comin}
is compared with calculated PDOSs
in the DFT+$U$ ($U=2.75$ eV) and the DFT+DMFT ($T=300$ K).
Blue dotted and
red solid lines represent the DFT+$U$ and the DFT+DMFT
PDOSs, respectively, which are broadened
with a Gaussian function (0.10 eV FWHM).
The Fermi level is adjusted
in the case of DFT+$U$ to fit the experiment.
Red solid and black dashed arrows point at
two dominating peaks in the DFT+DMFT and the measured PES, respectively.
Blue dotted arrows point at extra peaks in the
DFT+$U$, which become significantly broadened
in the DFT+DMFT due to the correlation-induced incoherency.
}
\label{fig2}
\end{figure}
%%%%%%%%%%%%%%%%%%%%%%%%%%%%%%%%%%%%%%%%%%%%%%%%%%%%%%%%%%%%%%%%
%%%%%%%%%%%%%%%%%%%%%%%%%%%%%%%%%%%%%%%%%%%%%%%%%%%%%%%%%%%%%%%%
\begin{figure*}[t]
\includegraphics[width=14cm]{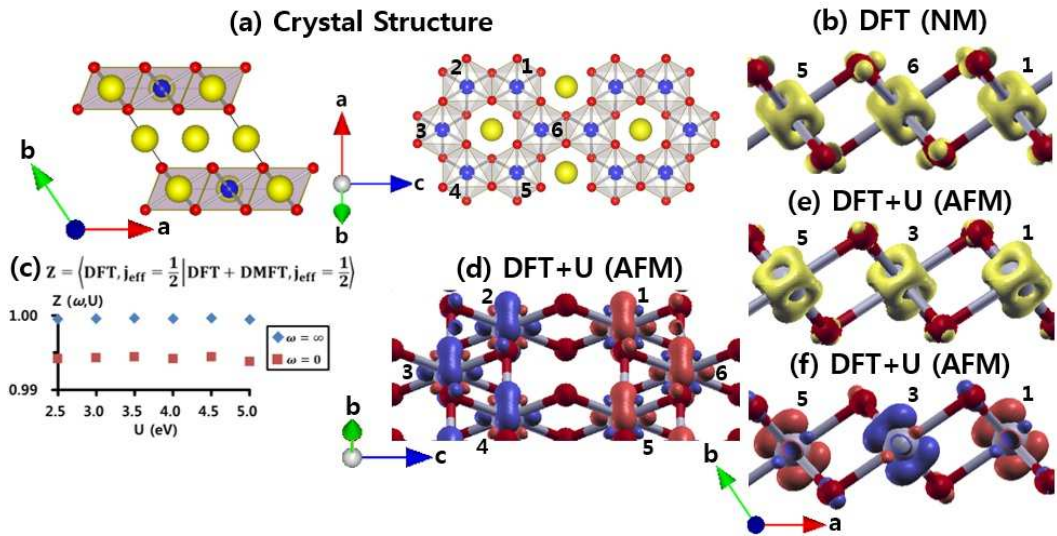}
\caption{(Color online)
(a) Crystal structure of Na$_{2}$IrO$_{3}$
(side view in the left and top view in the right).\cite{SKChoi}
Blue, yellow, and red balls correspond to Ir, Na, and O atoms,
respectively.
Numbers (1$-$6) in the top view are given to Ir atoms,
for which spin and charge densities are plotted in (b), (d), (e)-(f).
(b) The charge density plot of the unoccupied Ir t$_{2g}$ band
of the nonmagnetic (NM) state in the DFT.
Red balls represent oxygen atoms.
(c) Projection ($Z$) between
numerical $j_{eff}$=1/2 bases in the DFT and the DFT+DMFT
at $\omega$=0, and $\omega$=$\infty$ with variation of $U$.
(d) The spin density plot of zigzag-type AFM state in the DFT+$U$ ($U=2.75$ eV).
The obtained spin up (blue) and down (red) densities, respectively,
for 2,3,4 and 1,6,5 Ir atoms in the hexagon are
consistent with the zigzag-type AFM state.
(e) The charge density plot of
the unoccupied Ir t$_{2g}$ band of zigzag-type AFM state in the DFT+$U$ ($U=2.75$ eV).
(f) The spin density plot of zigzag-type AFM state in the DFT+$U$ ($U=2.75$ eV)
(different view from (d)).
Isosurfaces are chosen as 0.020 (e/a.u.$^{3}$)
for charge densities,
and 0.015 (e/a.u.$^{3}$) for spin densities.
}
\label{fig3}
\end{figure*}
%%%%%%%%%%%%%%%%%%%%%%%%%%%%%%%%%%%%%%%%%%%%%%%%%%%%%%%%%%%%%%%%

Figure \ref{fig1}(a) and (b) presents partial density of states (PDOS)
of $j_{eff}$=1/2, $j_{eff}$=3/2, and e$_{g}$ states in the DFT
and the DFT+DMFT, respectively.
The PDOS in the DFT in Fig. 1(a) shows six peaks in the t$_{2g}$ part.
Due to anisotropic solid environment in Na$_{2}$IrO$_{3}$,
the effective orbital degeneracy in the t$_{2g}$ manifold is seen to be
well lifted.
Mazin {\it et al.}\cite{Mazin12} interpreted these six peaks
as the QMOs resulting from strong intra- and weak inter-hexagon hybridizations.
On the other hand, in the two uppermost t$_{2g}$-driven bands,
$j_{eff}$=1/2 orbital (red solid) has main contribution with a small amount
of $j_{eff}$=3/2 (black dot) contribution. These QMO-driven six peaks
and dominant contribution from $j_{eff}$=1/2 near $E_F$
are consistent with previous experimental and theoretical
reports.\cite{CHSohn13,HSKim13,Foyevtsova}
The metallic PDOS in the DFT is not consistent with the insulating
ground state of Na$_{2}$IrO$_{3}$.

The PDOS in the DFT+DMFT in Fig. \ref{fig1}(b)
shows clear Mott gap of $\sim$400 meV size in the $j_{eff}$=1/2 states near $E_F$,
as is consistent with experiment.\cite{Comin}
Note that the DFT+DMFT calculation was done for the PM state (not AFM state)
at $T=300$ K well above $T_{N}$=15 K.
Therefore, the gap opens not due to the AFM cell-doubling
but due to the Coulomb correlation effect.
This result demonstrates that Na$_{2}$IrO$_{3}$ can be classified into
a Mott-type insulator.
This Mott insulating state of Na$_{2}$IrO$_{3}$ well above $T_{N}$
is different from the suggested marginal Mott insulating state
for Sr$_{2}$IrO$_{4}$.\cite{Zhang}
It is also noteworthy in Fig. \ref{fig1}(b) that,
due to the correlation-induced imaginary part
of the dynamical self-energy, t$_{2g}$-driven PDOSs in the DFT+DMFT
are significantly broadened with respect to those in the DFT.

The correlation-induced band gap and incoherent features
are more clearly seen in the spectral function plot in Fig. \ref{fig1}(c).
As shown in the Supplemental Material,\cite{Supp}
the spectral function becomes more and more incoherent with increasing $U$,
which clearly indicates that the incoherent feature is induced
by the correlation effect. It is also notable that the incoherent feature is
more prominent for $j_{eff}$=1/2 than for $j_{eff}$=3/2 states.

The correlation-induced incoherence
in the DFT+DMFT is essential to describe the measured PES of Na$_{2}$IrO$_{3}$.
In Fig. \ref{fig2}, Ir 5$d$ dominant PES near $E_F$
is compared with Ir 5$d$ PDOS obtained from both the DFT+$U$ and the DFT+DMFT.
In the PES, there seem to be two dominating peaks, which are
significantly broadened.\cite{Comin}
Our DFT+DMFT result shows the pronounced two peaks that are
broadened by the correlation-induced incoherence.
This dynamical correlation-induced incoherence feature is consistent
with previous report on Na$_{2}$IrO$_{3}$,\cite{Trousselet}
which analyzed the PES spectra based on the model Hamiltonian.
It is also shown in Fig. \ref{fig2} that, the DFT+$U$ with static correlation
yields four peaks, which is not consistent with the measured PES.

When describing the magnetic interaction in the
Mott insulating state of Na$_{2}$IrO$_{3}$,
the charge distribution of t$_{2g}$-driven
unoccupied state is usually considered as a rigid object,
which is determined by local interactions such as the
crystal field and the SOC.\cite{Jackeli,Chaloupka,Yamaji}
We have confirmed that onset of the zigzag-type AFM ordering
induces the significant redistribution of the unoccupied Ir t$_{2g}$
state, while the local dynamical correlation in the PM
state does not induce the redistribution.
Figure \ref{fig3} illustrates this argument.
In Fig. \ref{fig3}(b), the charge density of
the unoccupied Ir t$_{2g}$ band in the DFT is plotted.
The cubic-shaped charge density indicates the dominant $J_{eff}$=1/2
character of the band.
Figure \ref{fig3}(c) presents the projection $Z(\omega,U)$
between numerical $j_{eff}$=1/2 bases in the DFT and
the DFT+DMFT with varying $U$.
$j_{eff}$=1/2 bases were obtained from the diagonalization
of the hybridization function at zero and infinite frequency
($\Delta(0)$ and $\Delta(\infty)$), with and without considering
the self-energy contribution for the DFT+DMFT and the DFT, respectively.
$j_{eff}$=1/2 orbital for $\omega=\infty$ is relevant to
the single ion anisotropy, while $j_{eff}$=1/2 orbital for $\omega=0$
is relevant to the low energy electronic excitation.\cite{Zhang,Birol}
As shown in Fig.\ref{fig3}(c), $Z(\omega,U)$ is nearly
constant with respect to the variation of $U$, and
close to 1 for $\omega=\infty$ and 0.995 for $\omega=0$.
This result implies that numerical $j_{eff}$=1/2 bases
of the DFT and the DFT+DMFT resemble each other.
Namely, the local correlation does not affect the charge distribution,
and so the DFT+DMFT has similar charge distribution
of unoccupied t$_{2g}$-driven band
to that of the DFT.

On the other hand, the onset of the zigzag-type AFM ordering
induces significant elongation of charge density
along the ferromagnetic chain direction,
as shown in Fig. \ref{fig3}(e) for the DFT+$U$ result.
Spin densities in Fig. \ref{fig3}(d) and (f) also verify
this phenomenon. In the $J_{eff}$=1/2 basis, the spin
densities from three different orbitals in the t$_{2g}$
manifold have equal up, up, and down spin contributions.
In our DFT+$U$ results with the zigzag-type AFM ordering,
the orbital component directed normal
to the ferromagnetic chain is much smaller than
other two orbital components directed along the ferromagnetic chain.
As a result, the ratio between orbital and spin moment
($\mu_{L}/\mu_{S}$=1.59 with $\mu_{tot}$=0.70 $\mu_{B}$) deviates
largely from the ideal $J_{eff}$=1/2 case of $\mu_{L}/\mu_{S}$=2.
This feature indicates that there will be significant redistribution
of charge in the AFM transition due to the extended nature
of Ir 5$d$ orbital,\cite{Isosurface}
which is not expected for the superexchange-based magnetic interaction
in the normal Mott insulating state.

%%%%%%%%%%%%%%%%%%%%%%%%%%%%%%%%%%%%%%%%%%%%%%%%%%%%%%%%%%%%%%%%
\begin{figure}[t]
\includegraphics[width=8cm]{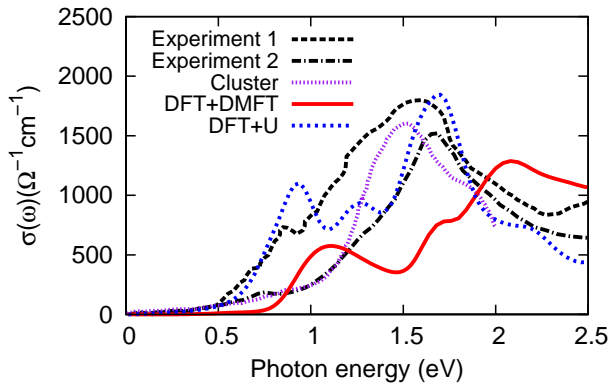}
\caption{(Color online)
Experimental optical conductivities $\sigma(\omega)$'s of Na$_{2}$IrO$_{3}$ at $T=300$ K,
Exp 1 from Ref. \cite{Comin} and Exp 2 from Ref.\cite{CHSohn13}, are compared
with calculated optical conductivities
using the DFT+DMFT, the DFT+$U$ ($U=2.75$ eV),
and four-site cluster multiplet calculations.\cite{BHKim14}
}
\label{fig4}
\end{figure}
%%%%%%%%%%%%%%%%%%%%%%%%%%%%%%%%%%%%%%%%%%%%%%%%%%%%%%%%%%%%%%%%

In Fig. \ref{fig4}, experimental optical conductivities $\sigma(\omega)$'s
are compared with computed optical conductivities from
the DFT+DMFT, the DFT+$U$,
and four-site cluster multiplet calculations.\cite{BHKim14}
In experiments,
only one peak is dominant in the optical spectrum.\cite{CHSohn13,Comin}
In the DFT+$U$, there occur two main peaks,
as in previous DFT+$U$ calculation,\cite{Li}
which does not seem to be consistent with experiments.
The peak near 0.75 eV should be suppressed.
In the DFT+DMFT, the peak near 0.75 eV is significantly
suppressed due to the incoherency from the local self-energy.
However, overall peaks are shifted upward
and the main peak position ($\sim$2 eV)
is not consistent with the experimental main peak position ($\sim$1.5 eV).
If we use smaller $U$ in the DFT+DMFT, the main peak position
is matched but the peak near 0.75 eV is not suppressed, and so
the resulting $\sigma(\omega)$ becomes similar to that of the DFT+$U$
(see the SM).\cite{Supp}
The inconsistent DFT+DMFT result suggests that the local dynamical
correlation is not sufficient to describe optical conductivity
of Na$_{2}$IrO$_{3}$, which corresponds to a two-particle property.

In contrast, the optical conductivity from the four-site cluster multiplet
calculation seems to be well consistent with experiments,
suggesting the importance of the non-local correlation effect
in Na$_{2}$IrO$_{3}$.\cite{BHKim14}
Similar feature was found in cuprate too, for which
PES that is single particle property is well described by
the single-site DMFT, but optical conductivity that is two-particle property
is described only by the cluster DMFT calculation.\cite{AGo}
Thus, it is expected that the non-local correlation effect would be
important for optical conductivity due to
the extended nature of Ir 5$d$ orbital in Na$_{2}$IrO$_{3}$.
More systematic study of cluster-size dependency of the 
optical spectrum would be an interesting future subject.

In conclusion, we show that
Na$_{2}$IrO$_{3}$ is a Mott-type insulator.
Differently from Sr$_{2}$IrO$_{4}$, the Mott insulating state
persists in the PM state well above $T_{N}$.
The local correlation-induced incoherence in
this Mott insulating phase explains the measured PES spectrum of Na$_{2}$IrO$_{3}$ well.
Yet, due to the extended nature of $5d$ orbital,
the insulating nature of
Na$_{2}$IrO$_{3}$ is different from that of a standard Mott insulator.
The onset of AFM ordering induces significant redistribution of charge.
The analysis of optical conductivity suggests that
the non-local correlation effect also plays a role in Na$_{2}$IrO$_{3}$.
Therefore, our results indicate that Na$_{2}$IrO$_{3}$ has
the Mott-type insulating nature,
but the itineracy of the charge degree of freedom
and the non-local correlation effect should also
be taken into account to describe physical properties of this system
having localized and itinerant duality of 5$d$ electrons.

%\begin{acknowledgments}
Helpful discussions with C.-J. Kang are greatly appreciated.
This work was supported by the POSTECH BK21+ Physics Project,
Max-Plank POSTECH/KOREA Research Initiative,
and the KISTI supercomputing center (No. KSC-2014-C3-044).
%\end{acknowledgments}


\begin{thebibliography}{99}
\bibitem{Isosurface}
      Strongly distorted spin density from ideal $J_{eff}$=1/2
      orbital in Ref. \cite{HJKim14} is similar to our results
      in Fig. \ref{fig3}(d) and (f).
      Our choice of isosurface electron density is nearly 25 times
      larger than that in  Ref. \cite{HJKim14}, and so
      our isosurface is less sensitive to the redistribution of charge
	  than that in Ref.\cite{HJKim14}
      Thus, captured distortions of charge and
      spin densities in our results indicate significant
      redistribution of electron density with onset of the zigzag-type AFM ordering.
\end{thebibliography}
\end{document}